\documentclass[12pt]{iopart}

\usepackage{color}
\usepackage{graphicx}
\usepackage{amssymb}

\usepackage{hyperref} 

\begin{document}
\title[
Large deviations of the branching random walk]
 {Slower deviations  of the branching Brownian motion  and of branching random walks }

\author{Bernard Derrida$^1$ and  Zhan Shi$^2$,}

\address{$^1$ Coll\`ege de France, 11 place Marcelin Berthelot, F-75231 Paris Cedex 05, France, and Laboratoire de Physique Statistique, \'Ecole Normale Sup\'erieure, Universit\'e Pierre et Marie Curie, Universit\'e Denis Diderot, CNRS, 24 rue Lhomond, F-75231 Paris Cedex 05, France, {\tt derrida@lps.ens.fr}}
\address{$^2$ Laboratoire de Probabilit\'es et Mod\`eles Al\'eatoires (LPMA), Universit\'e Pierre et Marie Curie, 4 place Jussieu, F-75252 Paris Cedex 05, France, {\tt zhan.shi@upmc.fr}}
\ead{derrida@lps.ens.fr,zhan.shi@upmc.fr}

\begin{abstract}
We have shown recently how to calculate the large deviation function of the position $X_{\max}(t) $ of the right most particle of a branching Brownian motion at time $t$.
This large deviation function exhibits a phase transition at a certain negative velocity. Here we extend this result to more general branching random walks and show that  the probability distribution of $X_{\max}(t)$ has, asymptotically in time,  a prefactor characterized by non trivial power law.
\end{abstract}
\ \\ \ \\
  \centerline{Dedicated to John Cardy on the occasion of his 70th birthday}
\ \\ \ \\
\submitto{\jpa}
\maketitle
\normalsize

\section{Introduction}
\begin{figure}[h]
\centerline{\includegraphics[width=10.cm]{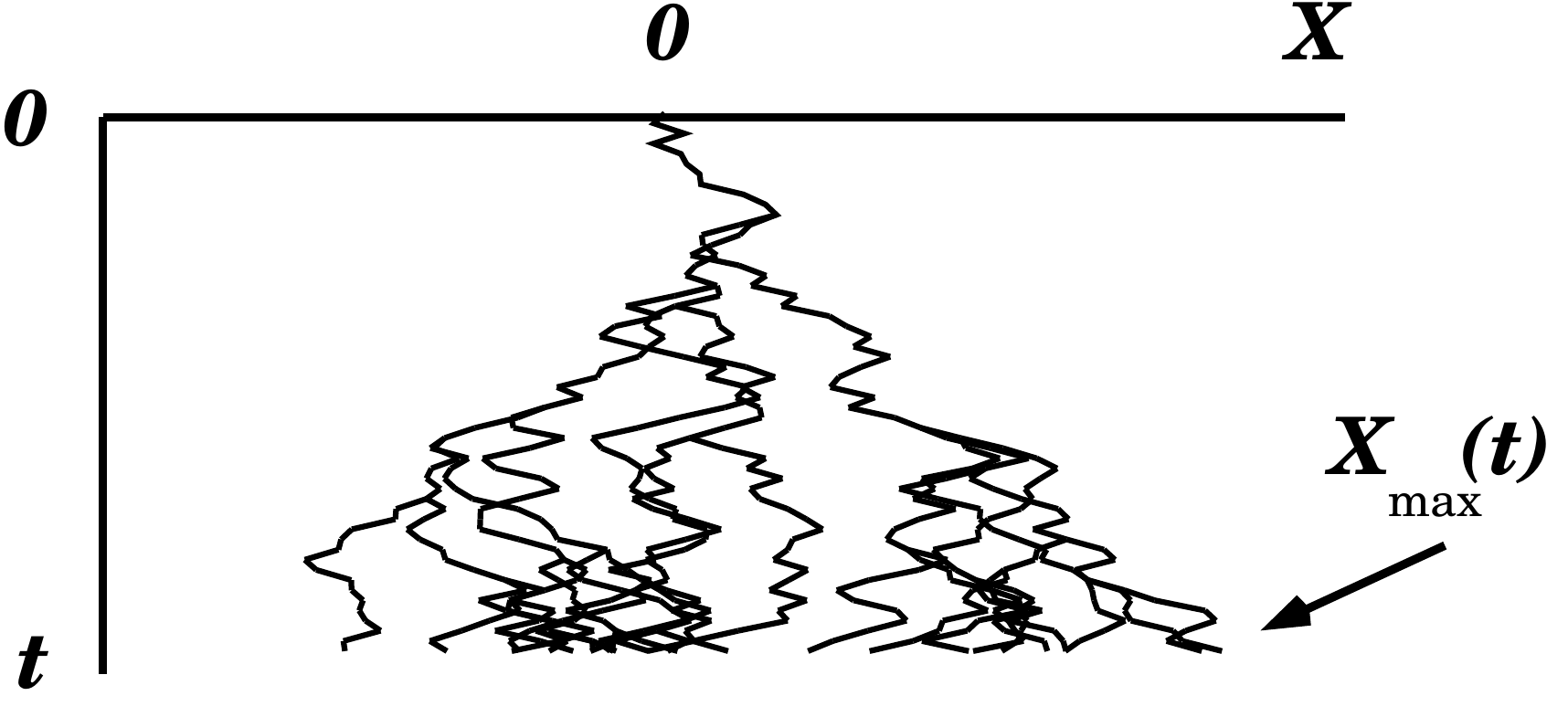}}
 \caption{A branching Brownian motion starts with a single particle at the origin. This particle performs a Brownian motion and branches at rate 1 into two independent branching Brownian motions  which themselves diffuse and branch, and so on. The number of particles grows typically like $e^t$ and we are interested in the position $X_{\max{}}(t) $ of the rightmost particle at time $t$.}
\label{figu}
\end{figure}

Branching Brownian motion and branching random walks  in one dimension have  been for several decades  a very active subject   among physicists  \cite{BD1,BD2,DSpohn,MK,MM,RMS}  and mathematicians \cite{ABBS,ABK,berestycki,Bovier,HS,LS,SK,Shi,zeitouni}  and mathematicians. One of the  most studied questions is how to determine  the distribution of the position  $X_{\max}(t)$ of the righmost particle of a branching Brownian motion at time $t$ (see figure \ref{figu}). 

 For a  branching Brownian motion, starting at the origin at $t=0$ as in figure \ref{figu}, the integrated distribution of the position $X_{\max}(t)$  defined by
\begin{equation}
u(x,t) = {\rm Prob }(X_{\max_{}} (t)  < x) 
\label{udef}
\end{equation}
 is known \cite{McKean} to satisfy   the Fisher-KPP equation:
\begin{equation}
{d u \over dt} = {d^2 \over dx^2} + u^2 - u \ \ \ \ \ {\rm with} \ \ \ \  u(x,0)=\theta(x) 
 \ . 
\label{KPP}
\end{equation}
(Here we take  the variance of the associated Brownian motion $\langle (X(t)- X(t'))^2 \rangle = \sigma^2 |t-t'| $ to be
   $\sigma^2=2$). 
In the long time limit the solution  becomes a travelling wave moving with an asymptotic velocity $2$.  More precisely one knows \cite{bramson78,bramson83}
 that in the long time limit, for $x-2 t = o(\sqrt{t})$,    
\begin{equation}
u(x,t) \sim F\left(x-2 t +{3\over 2} \log t -A \right) \label{tw}
\end{equation}
(where $X\sim Y$ means that that $ \lim \left({X \over Y}\right) \to 1$)
where the travelling wave $F(z)$ is  a solution of the differential equation 
\begin{equation}
\label{F-eq}
{d^2 F \over dz^2} + 2 {d F \over dz}  + F^2 - F=0
\end{equation}
such that
$F(z) \to 0 $ as $ z \to -\infty$
and $F(z) \to 1 $ as $ z \to \infty$.
Because of the shift invariance in (\ref{F-eq}),  if $F$ is a solution, $F$ translated along the $z$ axis is also a solution. To select  one particular   solution one can for example   specify  the  value  $F(0)$  at the origin, say $F(0)={1\over 2}$).
Once a particular solution $F$ of (\ref{F-eq}) has been chosen, the constant $A$ in (\ref{tw}) is fixed as well as the constant $B$ which is the prefactor
\begin{equation}
F(z) \sim B  e^{(\sqrt{2}-1) z} \ \ \ \ { \rm as} \ \ \ \ z \to - \infty \ . 
\label{FB}
\end{equation}
A  consequence  of (\ref{tw}) is that,  in probability,
\begin{equation}
{X_{\max{}} (t)  \over t} \to 2 \ \ \ \ \ {\rm as} \ \ \ \ \ t \to \infty
\ . 
\end{equation}

The question of large deviations \cite{DZ,DH} has to do with estimating  the long time asymptotics of $u(x,t) $ when  $x \sim c\,  t$ with $c \ne 2$  and to   calculate the large deviation function $\psi(c)$
defined by
\begin{equation}
{\rm Prob}(X_{\max{}} (t)  \sim  c \,  t )   \ \approx \ e^{ -t \, \psi(c) } 
\label{psi-def}
\end{equation}
(where $X \approx Y $  means that $\log X  \sim \log Y$).
For $c>2$ it is known \cite{chauvin-rouault,rouault} that $$
\psi(c)={c^2 \over 4} - 1$$
and that  even  the prefactor can be understood \cite{DMS}
$$1 - u(c\, t,t) \sim \left[ {1 \over \sqrt{4 \pi }}\left({2 \over c } - \int_0^\infty d \tau \int_{-\infty}^\infty dy \,  e^{-\left(1 + {c^2 \over 4} \right) \tau + {c  \over 2}y } \,  u^2(y,\tau)\right)   \right] {1 \over \sqrt{t}}  \   e^{- \left({c^2 \over 4} - 1 \right)t } $$
in terms of the short time solution of the F-KPP equation (\ref{KPP}).

In the present work we  estimate the long time asymptotics of $u(x,t) $ for
${x \over t} = c \ < 2.$ This large deviation function exhibits a phase transition \cite{DS}  at some negative velocity $c= 2 (1- \sqrt{2}) $ 
\begin{equation}
\psi(c) = \left\{ \begin{array}{lll}
(\sqrt{2}-1)(2-c) & {\rm for} & 2 (1- \sqrt{2}) < c < 2 \\
{c^2 \over 4} + 1  & {\rm for} & c < 2 (1- \sqrt{2})  
\end{array}
\right. \ . 
\label{res0}
\end{equation}
We will show in Section 2 that one can even understand the prefactor: for $2 (1- \sqrt{2}) < c < 2$
\begin{equation}
\label{res1}
u(c\, t,t) \sim
\left[ B \  e^{-A(\sqrt{2}-1)} \ \left(1-{2-c \over 2 \sqrt{2}} \right)^{3(\sqrt{2} -1) \over 2} \
 \right] 
 \ t^{3(\sqrt{2} -1) \over 2} \ 
 e^{t(\sqrt{2}-1)(c-2)}
\end{equation}
 where $A$ and $B$ are defined by (\ref{tw},\ref{F-eq},\ref{FB}) and that  for $c < 2 (1- \sqrt{2})$
\begin{equation}
\label{res2}
\fl  \qquad
 u(c \, t,t) \sim
\left[ {1 \over \sqrt{4 \pi }}\left(-{2 \over c } + \int_0^\infty d \tau \int_{-\infty}^\infty dy \, e^{\left(1 - {c^2 \over 4} \right) \tau + {c  \over 2} y}\,  u^2(y,\tau)\right)   \right] {1 \over \sqrt{t}}  \   e^{- \left({c^2 \over 4} + 1 \right) t }  \ . 
\end{equation}

Expressions (\ref{res1},\ref{res2}) are derived in Section 2.
Generalizations to a  family of   branching random walks and branching random motions   are discussed in Section 3.

\section{ The branching Brownian motion  }

It is easy to check that the solution of  the Fisher-KPP equation (\ref{KPP}) with the initial condition (\ref{udef}) can be written as 
\begin{equation}
u(x,t) =
 \int_{-\infty}^x   dy \ { e^{-t-{y^2 \over 4t}} 
 \over  \sqrt{4 \pi t }} 
\ + \ 
 \int_0^t d \tau \ \int_{-\infty}^\infty dy \   
{
 e^{-(t-\tau)  - {(x-y)^2 \over 4(t- \tau)}}
 \over  \sqrt{ 4 \pi (t-\tau) }}
 \  u^2(y,\tau) 
\ . \ \ \ 
\label{exact}
\end{equation}

In (\ref{exact}) the first term represents all the events  with no branching  up to time $t$ and while in the second term the integral over $\tau$ represents  the events for which the first branching event occurs at time $t-\tau$.

As we will see the reason for the phase transition between the two expressions in (\ref{res0}) is the following
\begin{itemize}
\item
 for $ 2 (1- \sqrt{2})< c <2  $   the second term in (\ref{exact}) dominates, and a saddle point calculation   will give that the optimmal $\tau$ is proportionnal to $t$.
\item
 for $c < 2 (1- \sqrt{2}) $ the two  
contributions in the r.h.s. of (\ref{exact}) are comparable, meaning that $u(x,t)$ is dominated by the events with no  branching at all or a first branching  at a very late time (i.e., $\tau \ll t $).
\end{itemize}
\ \\
One can   get a lower bound on $u(x,t)$  by keeping only the events such that there is no branching up to time $t-\tau$. Then for any $0 \le \tau \le t$
\begin{equation}
\int_{-\infty}^\infty  dy  {
 \ e^{-(t-\tau)  - {(x-y)^2 \over 4(t- \tau)}}
 \over  \sqrt{ 4 \pi (t-\tau) }}
 \  u(y,\tau)
\ \le \ u(x,t) \ .
\label{lower}
\end{equation}
Now the discussion is as follows:

Let us first assume    that the second term in (\ref{exact}) is  larger than the first term. Let $y_0$ and $\tau_0$  be  the  values of $y$ and $\tau$ which dominate the double integral in (\ref{exact}). (At this stage  we don't need to know the precise values of $y_0$ and $\tau_0$).
Then choosing $\tau=\tau_0$ in the lower bound (\ref{lower}), one gets
$$
\int_{-\infty}^\infty   dy \ 
 { e^{-(t-\tau_0)  - {(x-y)^2 \over 4(t- \tau_0)}} \over  \sqrt{ 4 \pi (t-\tau_0) }}
 \  u(y,\tau_0) 
< u(x,t) < 2 
 \int_0^t d \tau \ \int_{-\infty}^\infty    dy \ 
 { e^{-(t-\tau)  - {(x-y)^2 \over 4(t- \tau)}} \over  \sqrt{ 4 \pi (t-\tau) }}
  \  u^2(y,\tau) \  \ .   
$$
Because $u^2<u$, this double inequality can only be satisfied if    $u(y_0,\tau_0)$ is not  exponentially small in $t$: in other words,  $y_0 \gtrsim 2 \tau_0$. It is then easy to see that 

$$
 u(x,t) \ \approx \ \int_0^t d \tau \ \int_{2 \tau}^\infty  
dy \  { e^{-(t-\tau)  - {(x-y)^2 \over 4(t- \tau)}} \over  \sqrt{ 4 \pi (t-\tau) }}
$$
and the saddle point  over $\tau$ and $y$ is given by
\begin{equation}
y_0 = 2 \tau_0 \ \ \ \ \  {\rm and} \ \ \ \ \ \tau_0=  t \max \left[0, { c + 2 \sqrt{2} -2 \over 2 \sqrt{2} }\right]  \ . 
\label{saddle}
\end{equation}
We have therefore to distinguish two cases:
\begin{itemize}
\item If   $2 -2 \sqrt{2} < c < 2 $ (i.e., $\tau_0/t >0$ in (\ref{saddle}))  then the second term in (\ref{exact}) is exponentially (in the time $t$) larger  than the first term. In addition as $y_0 \sim  2 \tau_0$, the double integral in (\ref{exact}) is dominated by the region in $\tau$ and $y$ where one can use expression (\ref{tw}) for $u(y,\tau)$. Therefore
$$u(x,t) \sim    
 \int_0^t d \tau \ \int_{-\infty}^\infty 
dy \  { e^{-(t-\tau)  - {(x-y)^2 \over 4(t- \tau)}} \over  \sqrt{ 4 \pi (t-\tau) }}
 \   F^2\left(y-2 \tau +{3 \over 2} \log \tau  -A \right) \ . $$
Making the change of variable $z= y-2 \tau +{3 \over 2} \log \tau  -A $
and performing the integral over $\tau$ by a saddle point one gets
 \begin{equation}
\fl 
u(c\, t,t) \sim e^{t(\sqrt{2}-1)(c-2)} \
\left(   e^{-A} \, \left(1-{2-c \over 2 \sqrt{2}} \right)^{3 \over 2}\, t^{3 \over 2}  \right)^{\sqrt{2}-1}
 \, {1 \over \sqrt{8}}  \int dz \,  e^{z(1-\sqrt{2})} \, F^2(z)   \ . 
\label{res1bis}
\end{equation}

Finally one can show  (see the Appendix)  from
 (\ref{F-eq})  that
\begin{equation}
\int dz   \, e^{(1-\sqrt{2})z} F^2(z) =   \sqrt{8} \,  B
\label{FB1}
\end{equation}
and this leads  to 
$$ u( c\, t,t) \sim e^{t(\sqrt{2}-1)(c-2)} \ t^{3(\sqrt{2} -1) \over 2} \ e^{-A(\sqrt{2}-1)}  \ \left(1-{2-c \over 2 \sqrt{2}} \right)^{3(\sqrt{2} -1) \over 2} \      B $$
as announced in  (\ref{res1}).

\ \ \\ {\it Remarks:}
\begin{enumerate}
\item
 The values of $A$ and $B$ defined by (\ref{tw},\ref{FB}) depend on our choice of the solution $F$ of (\ref{F-eq}). One can see from (\ref{tw}) and (\ref{FB}) that
the combination $B \exp[-(\sqrt{2}-1) A ] $ which appears in (\ref{res1}) does not depend on any particular choice of $F$. 
\\ \ 
\item
One can also notice that when $c \to 2$ expression (\ref{res1})  reduces to (\ref{tw},\ref{FB}).
\end{enumerate}
\ \\ \

\item 
If $c<2 -2 \sqrt{2}  $  (i.e.,
$\tau_0/t=0$ in (\ref{saddle}))  the two terms in (\ref{exact}) are comparable and the
integrals are dominated by $\tau \sim y  =O(1)  \ll t$ and one obtains for large $t$
 expression (\ref{res2}).
\end{itemize}

\section{ A more 	 general branching random walk}
In this section we describe  how the above results (\ref{res0},\ref{res1},\ref{res2})
can be generalized to branching random walks. We  consider the following continuous time branching random walk in one dimension. One starts at $t=0$ with a single particle at the origin. This particle   jumps and branches at random times and when it branches, it gives rise to  $k$ independent new  branching random walks with probability $p_k$.  To be more precise,   if the number of particles is $N_t$ at time $t$, each particle (independently of what the other particles do)  branches with probability  $p_k dt $ into $k$ particles  and   moves a distance  between $y$  and $y+dy$ with probability $\rho(y) dy dt$ during every infinitesimal time interval $dt$.
\subsection{ The F-KPP like equation }
From this definition of the branching random walk, one can see that $u(x,t)$ defined as before by (\ref{udef}) satisfies
\begin{equation}
{d u(x,t) \over dt} = \int  \rho(y) dy \, 
 \big[u(x-y,t) - u(x,t) \big]
 \ + \  \sum_{k \ge  2} p_k \,  [u(x,t)^k- u(x,t) ]
\ \ \ 
\label{KPP2}
\end{equation}
with $u(x,0)=\theta(x)$.
As for the Fisher-KPP equation (\ref{KPP}) the uniform solutions $u=0$ and $u=1$ are respectively   stable and unstable and the solution of (\ref{KPP2}) with the initial condition $u(x,0)=\theta(x)$  becomes in the long time limit a travelling wave
\begin{equation}
u(x,t) \sim F\left(x-v_c \, t +{3\over 2 \gamma_c} \log t -A \right) \label{tw2}
\end{equation}
where $v_c$ and $\gamma_c$
are solutions of 
\begin{equation}
V'(\gamma_c) =0 \ \ \ \ \ ; \ \ \ \ \ v_c= V(\gamma_c) 
\label{vc}
\end{equation}
 where
\begin{equation}
V(\gamma)={1 \over \gamma} \left( \sum_{k \ge 2} (k-1) p_k \ + \ \int (e^{\gamma\,  y} - 1) \rho(y) d y \right) 
\label{v(gamma)}
\end{equation}
and the travelling wave $F(z)$ is a solution of
\begin{equation}
\int
 \, \rho(y)   d y
\, 
\big[F(z-y) - F(z) \big]
 \ +\ v_c \,  F'(z) \ + \ \sum_{k \ge  2} p_k [F(z)^k- F(z) ] = 0 
\label{F-eq2}
 \end{equation}
such that $F(-\infty) = 0$ and $F(\infty)=1$.

Expressions (\ref{tw2}),  (\ref{vc}) and (\ref{v(gamma)}) can be understood from (\ref{KPP2})  as follows:
if one looks for a travelling wave solution $u(x,t)=F(x-vt)$ of (\ref{KPP2}) moving at some velocity $v$, and one assumes that  $1-F(z) \approx e^{-\gamma z}$ close to the unstable solution (i.e., as  $z \to \infty$),  one gets by inserting this ansatz into (\ref{KPP2}) that $v=V(\gamma)$ with $V(\gamma)$ given by (\ref{v(gamma)}). Then as for all travelling wave  equations in the pulled case \cite{Van-Saarloos,BD3} one knows that  for steep enough initial conditions (in particular for $u(x,0)=\theta(x)$) the solution of (\ref{KPP2}) becomes a travelling wave solution of ({\ref{F-eq2}) moving at the velocity  $v_c$  given by (\ref{vc}) (i.e., the minimal velocity $V(\gamma)$ is selected) with a logarithmic shift in the position as in  (\ref{tw2}).

As for the branching Brownian motion, due to the shift invariance along the $z$ axis, one needs to select a solution of (\ref{F-eq2}) (for example by specifying  the value $F(0)$) and  this  fixes the value of $A$ 
in (\ref{tw2}) as well as the prefactor $B$ in the asymptotics of $F$ at $-\infty$
\begin{equation}
F(z) \ \sim \ B \, e^{\eta z}
\label{FB5}
\end{equation}
where $\eta$
is the positive solution of
\begin{equation}
\label{eta-eq}
\int \rho(y)dy \, \big( e^{-\eta y} -1 \big)         + \eta  \, v_c - \sum_{k\ge 2} p_k  \ =  \ 0 
\end{equation}
as can be easily checked by removing the non linear terms in (\ref{F-eq2}).

\subsection{The associated random walk}
Let us now  consider   the following  continuous time random walk (this random walk has the same stochastic evolution as the above branching random walk except that it does not branch). 

The probability $P(x,t)$ of finding this random walk  at position $x$ at time $t$ (if it starts at the origin) evolves according to
\begin{equation}
{d P (x,t)\over dt}  = \int \rho(y) d y \,  \big[ P(x-y,t) -P(x,t) \big] \ . 
\label{Prw}
\end{equation}
It is then easy to show that the generating function of $P(x,t)$ is given by
\begin{equation}
\label{ldp}
\int dx \, P(x,t) \,  e^{\gamma x} \ =  \  \exp \left[ t  \, 
g(\gamma) \right]  \ \ \ \ \ {\rm with} \ \ \ \ \
 g(\gamma) = \int \rho(y) dy \, \left(e^{\gamma\,  y} - 1\right)   \ . 
\end{equation}
From this exact expression, one can easily obtain (by a Legendre transform) the large deviation function $f(v)$ of the position of this random walk in a parametric form
\begin{equation}
f(v)= \gamma g'(\gamma)-g(\gamma) \ \ \ \ \ ; \ \ \ \ \ v=g'(\gamma)
\label{parametric}
\end{equation}
and even determine  the prefactor
\begin{equation}
P( x=v t,t) \sim  \sqrt{f''(v) \over 2 \pi t} \, e^{-t f(v)} \ .
\label{Pld}
\end{equation}

\subsection{The  large deviation function of $X_{\max_{}}(t)$ in the intermediate regime }
We are now going to  obtain the expressions which generalize the results (\ref{res0},\ref{res1},\ref{res2}) to the case of branching random walks.
As for  (\ref{exact}) one can check that
\begin{equation}
\fl
u(x,t)=
 e^{-\alpha t} 
 \int_{-\infty}^x dy \, P(y,t)  \, + \, \int_0^t d \tau 
\,  e^{-\alpha (t-\tau)} 
\int_{-\infty}^\infty dy \, P(x-y,t-\tau) \sum_{k \ge 2} p_k \, u^k(y,\tau) 
\label{u-general}
\end{equation}
with
\begin{equation}
\alpha= \sum_{k\ge 2} p_k 
\label{alpha-def}
\end{equation}
is solution of (\ref{KPP2}). As for the branching Brownian motion the first term represents the events with no branching up to time $t$ and, in the second term,  $t-\tau$ represents the time of the first branching event.

\ \\ \ \\
In the intermediate regime (i.e., when the second term in (\ref{u-general})
is exponentially larger than the first term),  one can show as we did for the branching Brownian motion 
that for the values of  $\tau$  and $y$ which dominate the second term on the r.h.s. of (\ref{u-general})  one can replace $u$ by its  expression (\ref{tw2}).
Therefore
writing $$y= v_c \, \tau - {3 \over 2 \gamma_c} \log \tau +A + z$$
and estimating the integral by the saddle point method over $\tau$ one gets
 after simplification
\begin{equation}
\fl u(c \,  t ,t) \sim \left[ \left({v_c-W \over c - W}\, t \right)^{-{3f'(W) \over 2 \gamma_c}}  \, {1\over v_c-W}  \,\int dz\,  e^{f'(W)(A + z)} \,  \sum_{k \ge 2} p_k F^k(z) \right]  \, e^{-t (c-v_c) f'(W)} 
\label{uc-general}
\end{equation}
where 
\begin{equation}
W= {c\,  t - v_c \, \tau_0 \over t - \tau_0}
\label{W-def}
\end{equation}
and $\tau_0$ is the saddle point of the integral over $\tau$.  By looking at the minimum over $\tau$
of $\alpha(t-\tau) +(t-\tau )f \left( {v t - v_c \tau \over t - \tau }\right)$
one finds that
$W$ is solution of 
\begin{equation}
\alpha + f(W) +(v_c-W) f'(W)=0  \ . 
\label{eq-W}
\end{equation}

Expressions (\ref{uc-general},\ref{W-def},\ref{eq-W}) are the generalization of (\ref{res1bis}) to the branching random walk.
As for the branching Brownian motion one can show show that (\ref{uc-general})  can be  further simplified to become
\begin{equation}
u(c \,  t ,t) \sim \left[ \left({v_c-W \over c - W}\right)^{-{3f'(W) \over 2 \gamma_c}}  \,  B \  e^{f'(W)\, A }  \right] \,  t^{-{3f'(W)  \over 2 \gamma_c}} \, e^{-t (c-v_c) f'(W)}
\label{uc-res}
\end{equation}
by using the facts (shown in the Appendix)  
\begin{equation}
\label{identity}
  \int dz \sum_{k \ge2} p_k \, F^k(z) e^{ f'(W) \,  z  }
= (v_c-W) B  \ \ \ \ \ {\rm and} \ \ \ \ \   \eta= -f'(W)
\end{equation}
where $\eta$  is the exponential decay of $F(z)$  as $z \to \infty$ (see (\ref{FB5})).

Expression (\ref{uc-res}) is the generalization of (\ref{res1}) in the intermediate regime, where 
the second term in (\ref{u-general}) is much larger than the first one, meaning that the saddle point $\tau_0/t$ is strictly positive, i.e.,
   \begin{equation}
\label{interm}W < c < v_c  \ . 
   \end{equation}
 In this regime, the large deviation (see (\ref{uc-res})) is linear in $c$
\begin{equation}
\psi(c)= -(v_c-c) f'(W)
\label{psiW}
\end{equation}
(remember that $f'(W) <0$) and, in the prefactor, there  is a power law  of time with an exponent $-3 f'(W)/2$ which is model-dependent.

\ \ \\ {\it Remark:}
The above calculation  and therefore (\ref{uc-res}) is only valid if the saddle point  is  in the range $0 < \tau_0 < t$
i.e., (see (\ref{W-def})), if 
\begin{equation}
    W\  < \  c \ < \ v_c \ . 
\label{cW-range}
\end{equation}
As for the branching Brownian motion, (\ref{uc-res}) matches with (\ref{tw2},\ref{FB5}) in the limit  $c \to  v_c$ (see (\ref{identity})) for the relation between $\eta$ and $f'(W)$).

For $c<W$, the formula which generalizes (\ref{res2}) is obtained from the range  $\tau= O(1) \ll t$  in (\ref{u-general}) and one finds
\begin{equation}
\fl u(c t ,t) \simeq e^{-(\alpha +f(c)) t} \sqrt{f''(c) \over 2 \pi t} \left[ -{1 \over f'(c)} + \int_0^\infty d \tau  \int  dz  e^{\tau( f(c)-c f'(c)) + z f'(c)} \, \sum_{k \ge 2} p_k \, u^k(z,\tau) 
\right] \ . 
\label{res2-bis}
\end{equation}

\section{An example}
All the discussion of section 2 can be easily generalized to a branching Brownian motion where, at each branching event, the particle branches into $m$ particles instead of 2. The Fisher-KPP equation (\ref{KPP})  becomes then 
\begin{equation}
{d u \over dt} = {d^2 \over dx^2} + u^m - u \ \ \ \ \ {\rm with} \ \ \ \  u(x,0)=\theta(x) 
 \ . 
\label{KPP7}
\end{equation}
In the long time limit the solution is given  as in (\ref{tw}) by
\begin{equation}
u(x,t) \sim F\left(x-v_c t +{3\over 2 \gamma_c} \log t -A \right) \label{tw5}
\end{equation}
where
\begin{equation}
v_c= 2 \sqrt{m-1} \ ÷ : \ \ {\rm and} \ \ \ \ \ \gamma_c= \sqrt{m-1} 
\label{vcm}
\end{equation}
and
the travelling wave $F(z)$ is solution of 
\begin{equation}
\label{FFF-eq}
{d^2 F \over dz^2} + 2 \sqrt{m-1} \ {d F \over dz}  + F^m - F=0 \ . 
\end{equation}
This allows to see that for $z \to \infty$
$$F(z) \sim  B e^{(\sqrt{m} - \sqrt{m-1} ) \, z} \ . $$

In this example the large deviation $\psi(c)$ defined by 
(\ref{psi-def})
 is given by
\begin{equation}
\fl 
\psi(c) = \left\{ \begin{array}{lll}
m-1-{c^2 \over 2}
& {\rm for} & v_c  < c  \\
(\sqrt{m}-\sqrt{m-1})(2 \sqrt{m-1}-c) & {\rm for} & W < c < v_c  \\
{c^2 \over 4} + 1  & {\rm for} & c <   W
\end{array}
\right.
\end{equation}
where $v_c$ is given by (\ref{vcm})
and
\begin{equation}
W= -2(\sqrt{m}-\sqrt{m-1}) \ . 
\end{equation}
We see that, as $f(v)=v^2/4$, one has $f'(W)=W/2$  so that the result in the intermediate regime
$W<c<v_c$ is,  as expected, given by (\ref{psiW})
and  (\ref{uc-res}) leads to 
\begin{equation}
u(c \,  t ,t) \sim \left[ \left({v_c-W \over c - W}\right)^{{3\over2} \left(\sqrt{m \over m-1}-1 \right)}  \,  B \  e^{f'(W)\, A }  \right] \,  t^{{3\over2} \left(\sqrt{m \over m-1}-1 \right)} \, e^{-t\,  \psi(c)} \ .
\label{res3}
\end{equation}
This expression reduces to (\ref{res1}) in the case $m=2$. One can notice as claimed in Section 3 that the power law of time in the prefactor  is model dependent.
As in (\ref{res1}) and in (\ref{uc-res}) the values of $A$ and $B$ depend on our arbitrary choice of the solution of (\ref{FFF-eq}) but the combination $B e^{f'(W) A}$ does not.

\section{Conclusion}
In this work we have obtained expressions of the large deviation function of the position of the rightmost particle of a branching Brownian motion (\ref{res0},\ref{res1},\ref{res2},\ref{res3}) and of  branching random walks (\ref{uc-res},\ref{res2-bis}). In general one observes a phase transition at some velocity $W$. As discussed at the beginning of Section 2, on one side of the transition, the events which dominate are those for which there is no branching at all or the first branching event occurs at a very late time. On the other side of $W$, the first branching event occurs at an intermediate time $\tau$ where $\tau \approx t-\tau \approx t$. 
One noticeable result is also that in the intermediate  regime, there is a power law of time in the prefactor with an exponent which is model-dependent.

It would be interesting to generalize our results to branching Brownian motions or branching random walks in presence of selection (for example to the $L$-BBM,  the $N$-BBM, coalescing random walks \cite{bzgd}) or to the noisy version of the F-KPP equation \cite{BDMM3,DMS,MMQ,MulS,PL}.

It is important to realize that our results depend strongly on the fact that we start with a single particle and that in the branching process, each particle has a non zero probability of not branching which decays exponentially with time. If we had considered an initial condition with more than one particle  as in \cite{meerson-sasorov} or  branching random walks which branch deterministically at equally spaced times, the probability distribution of $X_{\max_{}}(t)$ would be very different.

\appendix
\setcounter{section}{1}
\section*{Appendix}
In this appendix we establish relations (\ref{FB1}) and (\ref{identity}).
\subsection{Derivation of (\ref{FB1})}
For the branching Brownian motion, 
starting from (\ref{F-eq}),  we know that $F(z) \to 1$ as $z \to \infty$ and
$F(z) \to B e^{(\sqrt{2}-1) z}$ as $z \to \infty$ (see (\ref{FB})).
Therefore for any $0 < \eta  < 2(\sqrt{2} - 1) $
one has
$$
\int_{-\infty}^\infty dz \, F^2(z) \, e^{-\eta z} \ = \ 
\int_{-\infty}^\infty dz \,  \left[
F(z)
- 2 {d F(z) \over dz}  
- {d^2 F(z) \over dz^2} 
 \right]
\, e^{-\eta z}
$$
where the integrals converge.
One can then transform the right hand side by using integrations by parts
$$
\int_{-\infty}^\infty dz \, F^2(z) \, e^{-\eta z} \ = \ 
\lim_{\Lambda \to - \infty} \left[ 
  (2 + \eta) F(\Lambda) e^{-\eta \Lambda } \ + \ {dF(\Lambda )\over dz} e^{-\eta \Lambda } \ + \ (1 - 2 \eta- \eta^2)
\int_{\Lambda}^\infty dz \,
F(z) 
\, e^{-\eta z}
\right] \ .
$$
Thus for  $\eta=\sqrt{2}-1$ 
$$
\int_{-\infty}^\infty dz \, F^2(z) \, e^{- \eta  z} \ = \ 
\lim_{\Lambda \to - \infty} \left[ 
  (\sqrt{2}+1) F(\Lambda) e^{-\eta \Lambda } \ + \ {dF(\Lambda )\over dz} e^{-\eta \Lambda } 
\right]  \  =   \ 2 \sqrt{2} \,  B 
$$
where we used  (\ref{FB}).

\subsection{Derivation of (\ref{identity})}
The derivation of  (\ref{identity}) is very similar.  One  starts from (\ref{F-eq2})
and  gets
  $$\int dz \sum_{k \ge2} p_k \, F^k(z) e^{ - \eta z  }
  =\int dz   \left[\alpha F(z) - v_c F'(z) -
\int
  \rho(y)   d  y
\, 
\big[F(z-y) - F(z) \big]
\right] e^{ - \eta z  } $$
(where $\eta$ is,  for the moment, arbitrary  in the range where the integrals converge).
Because the integral  over $z$ converges
one can put a lower bound $\Lambda$ in the integrals  (and take later the limit  $\Lambda \to -\infty$)
and one can write the right hand side as
\begin{eqnarray*}
&  \int_{\Lambda}^\infty dz  
\left[\alpha F(z) - v_c F'(z) -
\int
  \rho(y       )   d y       
\,
\big[F(z-y       ) - F(z) \big]
\right] \, e^{-\eta z} =
\\
& \ \ \ \ \ \ \ \left[\alpha -v_c \eta -\int \rho(y) dy \,  (e^{-\eta y       }-1)        \right]  \int_{\Lambda}^\infty dz \,  F(z) \, 
e^{ - \eta z  }   \\ & \ \ \ \ \ \ \  \ \ \ \ \ \  \ \ \ \ 
+  v_c e^{-\eta  \Lambda} F(\Lambda) - \int \rho(y       ) d y       \ e^{-\eta y       } \int_{\Lambda - y       }^{\Lambda} dz \, F(z) \, e^{-\eta z} \ .
\end{eqnarray*}

Now for the particular value of  $\eta$ defined in (\ref{FB5}) and which satisfies (\ref{eta-eq}) (see also (\ref{alpha-def}))
one gets 
$$
  \int dz \sum_{k \ge2} p_k \, F^k(z) e^{ - \eta z  } \  = \ 
\lim_{\Lambda \to - \infty} \left[
  v_c\, F(\Lambda) \,  e^{-\eta  \Lambda}  - \int \rho(y       ) d y       \ e^{-\eta y       } \int_{\Lambda - y       }^{\Lambda}dz \, F(z) \,  e^{-\eta z} 
\right]
$$
and taking  the limit $\Lambda \to -\infty$ this gives
\begin{equation}
  \int dz \sum_{k \ge2} p_k \, F^k(z) e^{ - \eta z  } \  = \ 
  \left[v_c -\int         \rho(y       ) dy \, e^{-\eta y       }  \,  y       
\right] B \ .
\label{deriv1}
\end{equation}

Using (\ref{parametric}) one can see that 
\begin{equation}
g(\gamma)= v f'(v) - f(v)  \ \ \ \ \ ; \ \ \ \ \ \gamma = f'(v) \ .
\label{parametric1}
\end{equation}
This allows to write the equation (\ref{eq-W})  satisfied by $W$ as
\begin{equation}
-g(f'(W)) + v_c \, f'(W) + \alpha  =0 \ .
\label{fW}
\end{equation}
One can also rewrite (\ref{eta-eq})  using (\ref{ldp}) as
\begin{equation}
g(-\eta) + v_c \, \eta - \alpha =0\ .
\label{feta}
\end{equation}

\noindent  As such, both $-\eta$ and $f'(W)$ are solutions of $-g(u) + v_c \, u + \alpha  =0$. We now argue that they are identical. The function $G(u) := -g(u) + v_c \, u + \alpha$ is (strictly) concave, with $G(0) = \alpha >0$. So if $G$ has two roots, the larger root is positive. Since $-\eta$ and $f'(W)$ are negative, neither can be the larger root: both of them are the smaller root, and are identical. This shows as announced in (\ref{identity}) that
\begin{equation}
\eta = - f'(W) \ .
\label{etaW}
\end{equation}

Lastly 
from (\ref{parametric}) and (\ref{parametric1}) one has
$$v=g'(f'(v))$$
and from the definition (\ref{ldp}) of $g(\gamma)$  one gets
$$v = \int \rho(y) dy \, e^{f'(v) y} \, y  \ . $$
This, together with (\ref{etaW}) completes the derivation of (\ref{identity}).
\ \\ \ \\

\end{document}